\documentclass{PoS}
\pdfoutput=1

\usepackage{amsmath,amssymb,bbm,slashed}

\newcommand{\braket}[1]{\left\langle #1 \right\rangle}
\newcommand{\R}{\mathbbm{R}}
\newcommand{\Ds}{\slashed{D}}
\renewcommand\epsilon\varepsilon
\renewcommand\phi\varphi
\DeclareMathOperator{\1}{\mathbbm{1}}
\DeclareMathOperator{\diag}{diag}
\DeclareMathOperator{\str}{Str}

\title{Finite-volume corrections to low-energy constants from the
  partially quenched effective theory}

\ShortTitle{Finite-volume corrections to LECs from the
  partially quenched effective theory}

\author{\speaker{Christoph Lehner} and Tilo Wettig\\
  Institute for Theoretical Physics, University of Regensburg, 93040
  Regensburg, Germany\\
  E-mail: \email{christoph.lehner@physik.uni-regensburg.de},
  \email{tilo.wettig@physik.uni-regensburg.de}}

\abstract{We calculate finite-volume corrections to the low-energy
  constants $\Sigma$ and $F$ in the epsilon-regime of QCD using
  partially quenched chiral perturbation theory in the supersymmetry
  formulation without a singlet particle.  We comment on how to
  minimize these corrections in lattice simulations of QCD.}

\FullConference{The XXVII International Symposium on Lattice Field Theory - LAT2009\\
  July 26-31 2009\\
  Peking University, Beijing, China}

\begin{document}

\section{Introduction}

The low-energy constants (LEC) appearing in the chiral effective
Lagrangian, which are of great phenomenological importance, can be
determined by fitting analytical results from chiral random matrix
theory (RMT) to lattice data for the eigenvalue spectrum of the Dirac
operator.  The lowest-order LECs are $\Sigma$ and $F$.  While $\Sigma$
can be determined rather easily, e.g., from the distribution of the
small Dirac eigenvalues, the extraction of $F$ is somewhat more
complicated and requires the inclusion of a suitable chemical
potential \cite{Damgaard:2005ys,Akemann:2006ru}.

Since lattice simulations are restricted to a finite volume, it is
important to take into account finite-volume corrections to the RMT
results, which can be obtained by going to next-to-leading order (NLO) in
the $\epsilon$-regime.
Recently, finite-volume corrections to the unquenched partition function of QCD
in the $\epsilon$-regime were obtained in
\cite{Damgaard:2007xg,Akemann:2008vp}.  However, in order to extract
the relevant eigenvalue correlation functions the partially quenched
partition function of QCD is needed.  A relatively simple method to
obtain the partially quenched theory is to introduce $n$ replicated
flavors in the unquenched theory and then to analytically continue in
the discrete number of quark flavors to zero.  This so-called replica
trick was first used in the theory of disordered systems
\cite{Edwards:1975zz}.  It is well known that the replica trick is
potentially problematic since the analytic continuation from an isolated
set of points is not uniquely defined.

In this contribution we choose to use an alternative way to obtain the
partially quenched theory that does not suffer from the potential
problems of the replica trick and can therefore be used to check and
extend previous results.  In addition to the sea quarks, we introduce
fermionic and bosonic valence quarks.  In nuclear physics and
condensed matter physics this method is known as the supersymmetry
method or Efetov method for quenched disorder \cite{Efetov:1983zz}.
In the context of QCD this idea was first used by Morel
\cite{Morel:1987xk}.  The effective low-energy theory of QCD with
$N_f+N_v$ quarks and $N_v$ bosonic quarks was developed by Bernard and
Golterman \cite{Bernard:1993sv} and by Sharpe and Shoresh
\cite{Sharpe:2001fh}.  In this work we use the effective theory
without a singlet particle as
discussed by Sharpe and Shoresh and consider it in a finite volume and
for small quark masses.  In order to access $F$ in addition to
$\Sigma$, we include an imaginary quark chemical potential $\mu$
\cite{Damgaard:2005ys,Akemann:2006ru}.  (A first exploratory lattice
study of this idea was performed in Ref.~\cite{DeGrand:2007tm}.)  We
compute the partition function at next-to-leading order in the
$\epsilon$-regime and thereby obtain finite-volume corrections of
order $1/\sqrt{V}$ to the partially quenched theory that translate
into finite-volume corrections to the LECs $\Sigma$ and $F$.  Our
results agree with previous results for the unquenched partition
function \cite{Damgaard:2007xg,Akemann:2008vp}.  The details of this
calculation are given in a separate publication \cite{Lehner:2009pz}.

\section{The effective theory}
We define QCD with $N_f+N_v$ quarks and $N_v$ bosonic quarks by the
partition function
\begin{align} \label{eqn:partqcd}
  Z = \int d[A] \: e^{-S_\text{YM}}
  \Biggl[\prod_{f=1}^{N_f} \det(\Ds+m_f)\Biggr]
  \Biggl[\prod_{i=1}^{N_v}\frac{\det(\Ds+m_{vi})}
  {\det(\Ds+m'_{vi})}\Biggr] \,,
\end{align}
where the integral is over all gauge fields $A$, $S_\text{YM}$ is the
Yang-Mills action, $\Ds$ is the Dirac operator, $m_1,\ldots,m_{N_f}$
are the masses of the sea quarks, $m_{v1},\ldots,m_{vN_v}$ are the
masses of the fermionic valence quarks, and $m'_{v1},\ldots,m'_{vN_v}$
are the masses of the bosonic valence quarks.  By setting the mass
$m_{vi}$ of a valence quark equal to the mass $m'_{vi}$ of the
corresponding bosonic quark, the ratio of determinants of this pair
cancels and the flavor $i$ is quenched.  The corresponding chiral
effective theory in the partially quenched case of $N_f>0$ is defined
by the Nambu-Goldstone (NG) manifold given by \cite{Sharpe:2001fh}
\begin{align}
  U(x) &= \exp\left[\frac{i \sqrt 2}{F} \xi(x)\right]
\end{align}
with
\begin{align}\label{eqn:xi}
  \xi(x) &=  \begin{pmatrix}
    \pi(x) & \bar \kappa^T(x) \\
    \kappa(x) & i\pi'(x)
  \end{pmatrix}
  + \frac{i \phi(x)}{\sqrt{(N_f+N_v) N_v N_f}}
  \begin{pmatrix}
    N_v \1_{N_f+N_v} & 0 \\
    0 & (N_f+N_v) \1_{N_v}
  \end{pmatrix},
\end{align}
where $\kappa(x)$ and $\bar\kappa(x)$ are independent $N_v \times
(N_f+N_v)$ matrices with elements in the Grassmann algebra,
$\pi(x)=\pi(x)^\dagger$ and $\pi'(x)=\pi'(x)^\dagger$ are traceless
Hermitian matrices of dimension $N_f+N_v$ and $N_v$, respectively,
$\phi(x) \in \R$, and $\1_n$ is the $n$-dimensional identity matrix.
Note that in the fully quenched case of $N_f=0$ also the singlet field
needs to be included in the effective theory \cite{Sharpe:2001fh}.  In
this contribution, however, we restrict the discussion to the
partially quenched case.  The Lagrangian of the effective theory to
leading order in $U(x)$, $\partial_\rho U(x)$, and the quark mass
matrix $M$ is given by
\begin{align}
  {\cal L} & = \frac{F^2}4 \str \left[ \partial_\rho U(x)^{-1} \partial_\rho U(x) \right] 
  - \frac{\Sigma}{2} \str\left[ M^\dagger U(x) + U(x)^{-1} M\right]
\end{align}
with $M=\diag(m_1,\ldots,m_{N_f}, m_{v1},\ldots,m_{vN_v},
m'_{v1},\ldots,m'_{vN_v})$ and low-energy constants $\Sigma$ and $F$.
The Lagrangian for nonzero imaginary chemical potential is obtained by
replacing the derivative $\partial_\rho$ with the covariant derivative
$\nabla_\rho$ defined by
\begin{align}\label{eqn:nablachem}
\nabla_\rho U(x) = \partial_\rho U(x) -i \delta_{\rho 0}[C,U(x)]\,,
\end{align}
where $C=\diag(\mu_1,\ldots,\mu_{N_f}, \mu_{v1},\ldots,\mu_{vN_v},
\mu'_{v1},\ldots,\mu'_{vN_v})$ and $i\mu_i$ is the imaginary chemical
potential of quark flavor $i$.  The supertrace $\str$ is a
generalization of the ordinary trace that satisfies $\str AB = \str
BA$ \cite{Efetov:1983zz}.

\section{Finite-volume corrections to $\Sigma$ and $F$}
In this section we consider the theory in a box of volume $V=L_0 L_1
L_2 L_3$ in the Euclidean formalism.  The temporal extent of the box
is given by $L_0$.  We use the $\varepsilon$-regime power counting
\cite{Gasser:1987ah} defined by
\begin{align}
  V \sim \varepsilon^{-4}\,,\qquad M \sim \varepsilon^4\,,\qquad \mu
  \sim \varepsilon^2\,,\qquad
  \partial_\rho \sim \varepsilon\,,\qquad \xi(x) \sim \varepsilon\,.
\end{align}
Note that the expansion in $\varepsilon^2$ amounts to an expansion in
$1/\sqrt{V}$.  To leading order in this power counting the fields
$\xi$ are effectively massless.  Therefore, in order to obtain finite
propagators, we separate the constant mode $U_0$ by the ansatz
\begin{align}\label{eqn:pu}
  U(x) &= U_0 \: \exp\left[\frac{i \sqrt 2}{F} \xi(x)\right]
\end{align}
with $\int d^4x\: \xi(x) = 0$.  The leading-order Lagrangian is given
by
\begin{align}
  {\cal L}_0 &= \frac{1}{2} \str \left[(\partial_\rho\xi(x))
    (\partial_\rho\xi(x))\right] -\frac\Sigma 2\str\left[ M^\dagger
    U_0 + U_0^{-1} M \right] -\frac{F^2}4\str \left[ [C, U_0^{-1}] [C,
    U_0] \right].
\end{align}
A careful analysis of the propagator, see Ref.~\cite{Lehner:2009pz},
yields
\begin{align}
  \braket{ \xi(x)_{ab} \xi(y)_{cd} }_0 &= \bar\Delta(x-y) \left[
    \delta_{ad}\delta_{bc} (-1)^{\varepsilon_b} - {\frac1{N_f}}\delta_{ab} \delta_{cd} \right],
\end{align}
where 
\begin{align}
  \braket{{\cal O}[\xi]}_0 = \frac{\int d[\xi]\: {\cal O}[\xi]\: e^{-\int d^4x\,
      {\cal L}_0}} {\int d[\xi]\: e^{-\int d^4x\:{\cal L}_0}}\,,
\end{align}
$\bar\Delta(x)$ is the massless propagator without zero modes
\cite{Hasenfratz:1989pk}, and
\begin{align}
  \varepsilon_b =
  \begin{cases}
    0 & \text{for } 1\le b \le N_f + N_v\,,\\
    1 & \text{for } N_f + N_v < b \le N_f + 2 N_v\,.
  \end{cases}
\end{align}
Note that the propagator does not depend on the valence quark number
$N_v$.  The propagator $\bar\Delta(x)$ is finite in dimensional
regularization and contains the dependence on the volume.
The contributions to the Lagrangian at next-to-leading order in
$\varepsilon$ are given by
\begin{align}
  {\cal L}_2 &= {\cal L}^M_2 + {\cal L}^C_2 + {\cal L}^N_2
\end{align}
with
\begin{align}
  {\cal L}^M_2 &=\frac{\Sigma}{2 F^2} \str\, \bigl[ M^\dagger U_0
    \xi(x)^2 + \xi(x)^2 U_0^{-1} M \bigr], \\
  {\cal L}^C_2 &= -\frac 12 \str  U_0^{-1} C U_0
    [\xi(x),[C,\xi(x)]] -\frac i2 \str\, (U_0^{-1} C
    U_0+C) [\xi(x), \partial_0\xi(x)]\,, \\
  {\cal L}^N_2 &= \frac{1}{12F^2}\str\,
    [\partial_\rho\xi(x),\xi(x)][\partial_\rho\xi(x),\xi(x)]
    -\frac1{3\sqrt2F} \str \, U_0^{-1} [C, U_0] [\xi(x),
    [\partial_0\xi(x), \xi(x)]] \,.
\end{align}
We will integrate out the fluctuations in $\xi$ in order to
obtain an effective finite-volume partition function.  The term ${\cal
  L}^M_2$ couples to $U_0$ and $M$ and corrects the leading-order mass
term to
\begin{align}
  -\frac\Sigma2 \left[1 - \frac{N_f^2-1}{N_f F^2}
    \bar\Delta(0)\right] \str\left[ M^\dagger U_0 + U_0^{-1} M
  \right],
\end{align}
where $\bar\Delta(0)$ in dimensional regularization is given by
\cite{Hasenfratz:1989pk}
\begin{align}
\bar\Delta(0) = -\frac{\beta_1}{\sqrt V}\,. 
\end{align}
The so-called shape coefficient $\beta_1$ only depends on the
quantities $l_i = L_i / V^{1/4}$ with $i=0,1,2,3$.  The terms in
${\cal L}^C_2$ couple to $U_0$ and $C$ and correct the leading-order
chemical potential term to
\begin{align}
  -\frac{F^2}4 \left[1 -\frac
      {2N_f}{F^2}\Bigl(\bar\Delta(0)-\int d^4x
        \left(\partial_0\bar\Delta(x)\right)^2\Bigr) \right]\str
  \left[ [C, U_0^{-1}] [C, U_0] \right]
\end{align}
with 
\begin{align}\label{eqn:knn}
\int d^4x \left(\partial_0\bar\Delta(x)\right)^2 = -\frac1{2\sqrt
V}\left[\beta_1-\frac{L_0^2}{\sqrt V} k_{00}\right],  
\end{align}
where $k_{00}$ is another shape
coefficient \cite{Hansen:1990un} that only depends on $l_0,\ldots,l_3$.
Thus we can read off effective low-energy constants
$\Sigma_\text{eff}$ and $F_\text{eff}$,
\begin{align}
  \frac{\Sigma_\text{eff}}{\Sigma} &= 1 - \frac{N_f^2-1}{N_f F^2}
  \bar\Delta(0)\,,\\
  \frac{F_\text{eff}}{F} &= 1 -\frac {N_f}{F^2}\Bigl(\bar\Delta(0)-\int
    d^4x \: \left(\partial_0\bar\Delta(x)\right)^2\Bigr)\,.
\end{align}
This is the same result as previously derived for the unquenched
theory \cite{Damgaard:2007xg,Akemann:2008vp}.  The terms in ${\cal
  L}^N_2$ and also potential contributions from the integration
measure are not relevant for finite-volume corrections to $\Sigma$ and
$F$ at next-to-leading order in $\varepsilon$, see
Ref.~\cite{Lehner:2009pz} for details.

\section{The constant-mode integral}
In the infinite-volume limit the fluctuations in $\xi$ are suppressed
and the theory becomes zero-dimensional.  It is therefore described by
RMT, and the integral over the constant mode $U_0$ defined in
Eq.~\eqref{eqn:pu} should recover the RMT result.
A proof for the case of vanishing chemical potential is given in
Ref.~\cite{Lehner:2009pz}.
We also note that fixing topology to a single topological sector $\nu$
merely amounts to a change of the integration manifold of $U_0$ and
therefore does not alter the discussion of the finite-volume
corrections given above.  For a detailed discussion of the theory in a
single topological sector we refer to Ref.~\cite{Lehner:2009pz}.

\section{Conclusions}
In this work we have calculated the partially quenched partition
function of QCD at next-to-leading order in the
$\varepsilon$-expansion at nonzero imaginary chemical potential.  We
considered a theory with $N_f+N_v$ fermionic quarks and $N_v$ bosonic
quarks, as formulated by Sharpe and Shoresh \cite{Sharpe:2001fh}, in a
finite volume $V$ with microscopic quark masses $M$, i.e., $M V \Sigma
= {\cal O}(\varepsilon^0)$.  The knowledge of the analytic form of the
partially quenched partition function suffices to obtain all spectral
correlation functions of the Dirac operator $\Ds$.  In this sense our
results for the finite-volume behavior of the theory hold universally
for all observables that can be obtained from spectral correlation
functions of $\Ds$.  We found that the partially quenched partition
function has the same finite-volume corrections as the unquenched
partition function of QCD with $N_f$ quarks.

\begin{figure}[t]
  \centering
  \includegraphics[width=7.4cm]{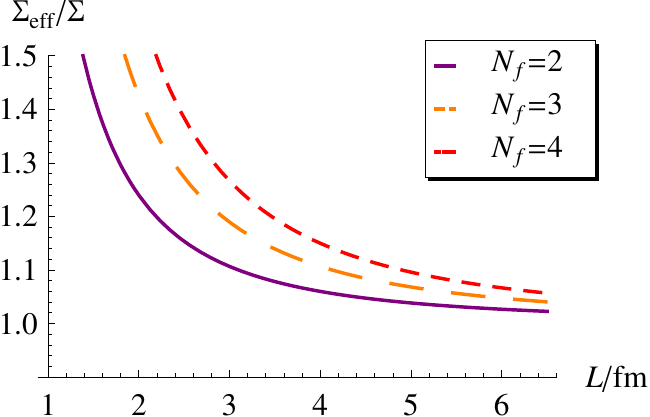}
  \includegraphics[width=7.4cm]{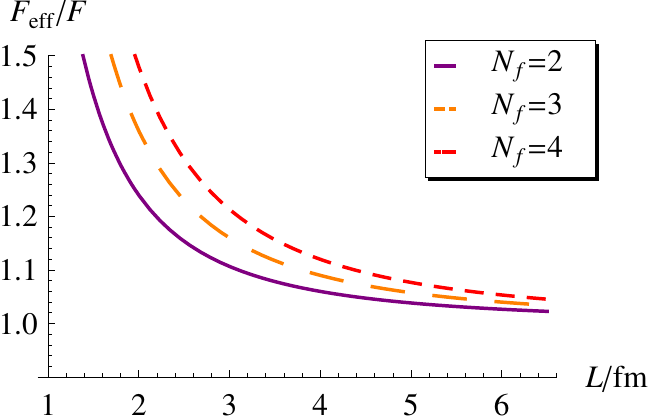}
  \caption{Volume dependence at NLO of the low-energy constants $\Sigma_\text{eff}$ (left)
    and $F_\text{eff}$ (right) in a symmetric box with dimensions $L_0=L_1=L_2=L_3=L$ at $F=90$ MeV.}
  \label{fig:p}
\end{figure}
\begin{figure}[t]
  \centering
  \includegraphics[width=7.3cm]{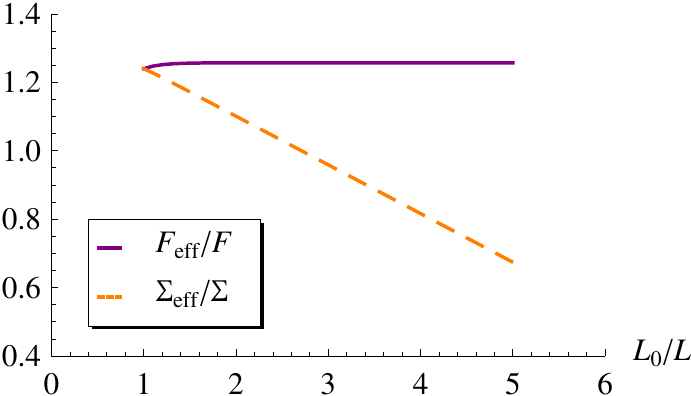}
  \hspace{0.2cm}
  \includegraphics[width=7.3cm]{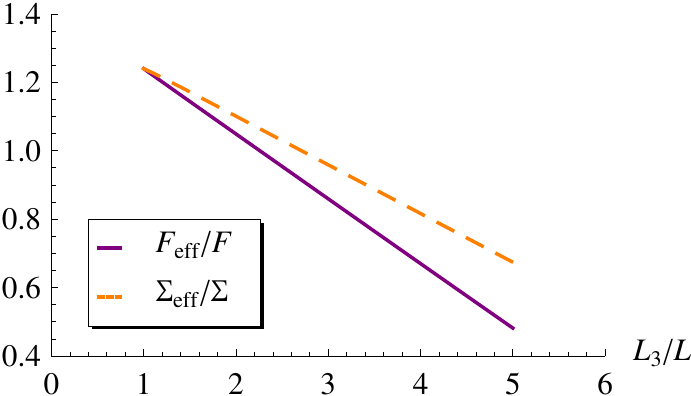}
  \caption{Effect of an asymmetric box with parameters $N_f=2$, $L=2$ fm, and $F=90$ MeV.
    We compare a large temporal dimension $L_0$ with $L_1=L_2=L_3=L$ (left) to
    a large spatial dimension $L_3$ with $L_0=L_1=L_2=L$ (right).
  }
  \label{fig:pratio}
\end{figure}
In Fig.~\ref{fig:p} we show the finite-volume corrections at NLO to
the low-energy constants $\Sigma$ and $F$ as a function of the box
size $L$ in a symmetric box.  Note that the effects of the finite
volume increase with the number of sea quark flavors $N_f$ and that,
depending on $N_f$, a box size of $3 - 5$ fm is necessary to reduce
the effects of the finite volume at NLO to about 10\%.  The effects
are calculated at $F=90$ MeV.  In Fig.~\ref{fig:pratio} we show the
effect of an asymmetric box with $N_f=2$ and $L=2$ fm.  An important
message of this figure is that the magnitude of the finite-volume
corrections can be significantly reduced by choosing one large spatial
dimension instead of a large temporal dimension.  The reason for this
behavior is that the chemical potential only affects the temporal
direction, see Eq.~\eqref{eqn:nablachem}, and therefore breaks the
permutation symmetry of the four dimensions.  This manifests itself in
the propagator
\begin{align}
  \int d^4x \left(\partial_0 \bar\Delta(x)\right)^2
\end{align}
which, as shown in Eq.~\eqref{eqn:knn},
contains a term proportional to $L_0^2 / \sqrt V$, where $L_0$ is the size of the temporal
dimension.  This term leads to an enhancement of the corrections in case of a large temporal
dimension.
Choosing instead one large spatial dimension, the finite-volume corrections are reduced,
unless the asymmetry is too large.  For the parameters used in
Fig.~\ref{fig:pratio}, the optimal value is $L_3/L \approx 2$.

This is good news.  Many lattice simulations (at zero chemical
potential) are performed with $L_1=L_2=L_3=L$ and $L_0=2L$.  To
determine $F$, it suffices to introduce the imaginary chemical
potential in the valence sector.  Therefore, one can take a suitable
set of existing dynamical configurations and redefine
$L_0\leftrightarrow L_3$ before adding the chemical
potential.\footnote{Note that this procedure increases the
  temperature of the system by a factor of two.  One needs to check
  that the system does not end up in the chirally restored phase, in
  which our results no longer apply.}  This will minimize the
finite-volume corrections for both $\Sigma$ and $F$, at least for the
parameter values chosen in Fig.~\ref{fig:pratio}.

\acknowledgments
We thank Hidenori Fukaya and Shoji Hashimoto for stimulating
discussions and the Theory Group of the INPS at KEK Tsukuba for their
hospitality.  This work was supported in part by BayEFG (CL) and by
DFG and KEK (TW).

\bibliographystyle{JHEP}
\bibliography{lattice09} 

\end{document}